\documentclass[NATO,numreferences]{crckbked}
\usepackage{epsfig}

\begin{opening}
\title{P-wave Pairing in superconducting Sr$_2$RuO$_4$}

\author{G. LITAK}
\institute{Department of Mechanics, Technical University of Lublin \\
Nadbystrzycka 36, Lublin PL--20--618, Poland}
\author{J.F. ANNETT, B.L. GY\"ORFFY}
\institute{H.H. Wills Physics Laboratory, University of Bristol \\
Tyndall Ave, Bristol BS8 1TL, United Kingdom }
\author{K.I. WYSOKI\'NSKI}
\institute{Institute of Physics, M. Curie--Sk\l odowska University, \\
Radziszewskiego 10a, Lublin PL--20--031, Poland }

\end{opening}

\begin{document}

\def\refname{7.~References}


\section{Introduction}

 Although there is strong  evidence that Sr$_{2}$RuO$_{4}$ is a
triplet
superconductor \cite{Mae01,Mac00}, 
the full symmetry of the equilibrium
state
below T$_{c}$ remains open to
debate [2-14]. 
There exist strong indications for broken time-reversal
symmetry in the superconducting state \cite{Luk98,Sig00}
 and equally convincing measurements
showing that the order parameter ${\bf d}({\bf k})$ has a
 line of nodes on the Fermi
surface [16-19].
The reason why this state of affairs represents a puzzle is
that for all odd parity spin triplet pairing states in tetragonal
crystals,
 group theory does not require the simultaneous presence of both broken
time-reversal symmetry and line nodes \cite{Ann90}.
To explain this inconsistency 
different three dimensional models of pairing have been proposed 
[11-13]. In fact the experimental results on heat
transport \cite{Tan01} seem to
favour 
the horizontal, with respect to (ab) crystal plane, line nodes.
Usually in a multi-band
BCS
like model, with different coupling constants for each band, one
generically
finds multiple phase transitions as the different sheets of the Fermi
surface are gaped on lowering the  temperature. Since experimentally
there
is
only one jump in the specific heat, at $T_{c}=1.5$ K, in constructing a
sensible model one must eliminate such multiple
transitions.  Zhitomirsky and Rice \cite{Zhi01}, in their simplified two
band
model, considered an interaction which couple the order parameters of the
different symmetry. The presence of this inter-band interaction leads to a
single superconducting transition.
Microscopically such coupling of different bands comes from effective
three
site interactions.  
 They assumed that one
band in Sr$_2$RuO$_4$ is the most important for superconducting pairing,
while two other are gaped via
an inter-band proximity effect. On the other hand Annett {\it et al.}
\cite{Ann01} have proposed sightly different model. They considered a three
orbital three dimensional model with effective in plane and out of plane
nearest neighbour interactions, but did not allow for
interactions mixing the symmetry of the order parameters. The
price to pay in this model is the fine-tuning of two interactions in order
to have a single superconducting transition at $T_c \approx 1.5$ K. The
calculations show that the model \cite{Ann01} explains quantitatively the
$T$ dependence of the specific heat, penetration depth and thermal
conductivity without additional fitting parameters. 

It is the purpose of
this work to extend the model of Annett {\em et al.} \cite{Ann01} by
allowing for symmetry mixing
interactions. To this end we 
add small three point interaction to describe possible
inter-orbital proximity effects.

\begin{figure}[thb]
\vspace*{-0.5cm}
\centerline{\epsfig{file=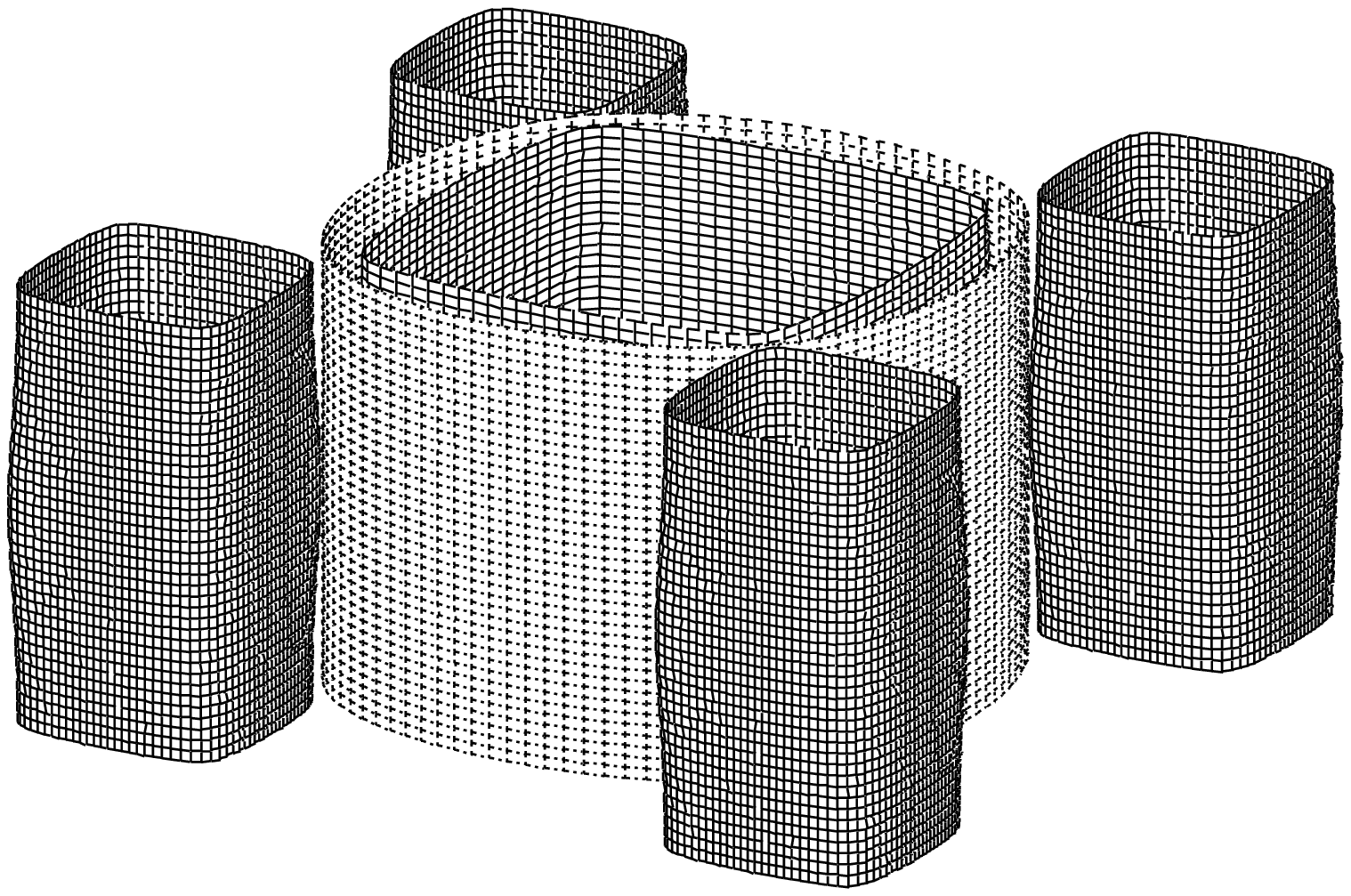,width=10.0cm,angle=-90}}
\vspace{-6.4cm}
\hspace{6.2cm}
\epsfig{file=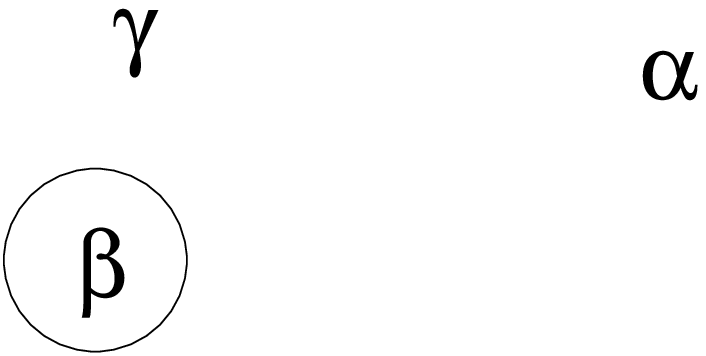,width=2.5cm,angle=0}
\vspace{3cm}
\caption{Calculated tight binding Fermi surface of Sr$_2$RuO$_4$.}
\label{fig1}
\end{figure}

\section{The orbital model of superconductivity}

To describe the superconducting state we use a simple multi-band
attractive Hubbard model. Its Hamiltonian consists of two parts:
\begin{equation}
\hat H=\hat H_0 + \hat H_{int},
\label{eq1}
\end{equation}
where $\hat H_0$ denotes the tight binding electron part corresponding to
the
experimentally observed band
structure of Sr$_2$RuO$_4$. 
\begin{equation}
  \hat{H_0} = \sum_{ijmm',\sigma}
\left( (\varepsilon_m  - \mu)\delta_{ij}\delta_{mm'}
 - t_{mm'}(ij) \right) \hat{c}^+_{im\sigma}\hat{c}_{jm'\sigma}, 
\label{eq2}
\end{equation}
here $m,m'=a,b,c$ refer to the three ruthenium $t_{2g}$ orbitals $a=xz$,
$b = yz$ and $c = xy$ and  $i$ and $j$ label the sites
of a body centered tetragonal lattice.
$c^{\dagger}_{i m
\sigma}$ and $c_{i m \sigma}$ are the Fermion
creation and annihilation operators for an electron
 on site $i$ and orbital $m$ with spin $\sigma$.

The hopping integrals $t_{mm'}(ij)$ and site
energies $\varepsilon_m$ were fitted to reproduce the experimentally
determined Fermi surface \cite{Ber00}. The calculated Fermi surface is
shown in Fig.~\ref{fig1}.

\begin{figure}[thb]   
\centerline{
\epsfig{file=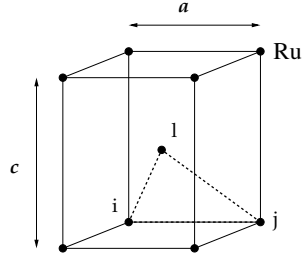,width=3.6cm,angle=0}}
\vspace{0cm}
\caption{Body centered tetragonal lattice of Ru atoms in Sr$_2$RuO$_4$. 
Sites $i$, $j$, $l$ correspond of possible realization
of three nearest lattice sites. Note that two 
of sites must lie in RuO$_2$ plane, while the third lies outside it.}
\label{fig2}
\end{figure}  

The second part $\hat H_{int}$ (Eq.~\ref{eq1}) describes the general 
electron--electron interaction and can be written as
\begin{equation}
H_{int}=  -\frac{1}{2} \sum_{ijlsmm'oo'\sigma\sigma'}
U_{mm'oo'}^{\sigma\sigma'}(ijls)
c^+_{im \sigma}c_{jm' \sigma}c^+_{lo \sigma'}c_{so' \sigma'}.
\label{eq3}
\end{equation}

Some of  the  interaction
constants $U_{mm'oo'}^{\sigma\sigma'}(ijls)$ are assumed to be attractive:
those acting
between electrons on nearest neighbour sites
with parallel spins 
and responsible for superconductivity.
In Eq. \ref{eq3}  all site indices $ijls$ can correspond, in
general, to
different sites but we assume, as
is usually the case in an isotropic substance, that $n$--point
interactions
satisfy the relation
\begin{equation}
|U^{(n=1)}|>|U^{(n=2)}|>|U^{(n=3)}|>|U^{(n=4)}|.
\label{eq4}
\end{equation}

A single point 
interaction ($U^{(1)}$)  refers, as usual, to the 
on-site repulsive Coulomb repulsion which
does not
contribute to
electron pairing while  the  $U^{(2)}$, assumed to be attractive,
represent
the main contribution.  
The three point interaction coupling ($U^{(3)}$)  is apparently
smaller than the bond term ($U^{(2)}$), however its influence on 
pairing mechanism may be important \cite{Zhi01}, so we retain it,
neglecting a four point coupling $U^{(4)}$. 

Thus, the interaction part of 
Hamiltonian in our case reads:
\begin{eqnarray}
\hat H_{int} &=&  -\frac{1}{2} \sum_{ijmm'\sigma\sigma'}
U_{mm'}^{\sigma\sigma'}(ij)
\hat n_{i m \sigma} \hat n_{j m' \sigma'} \nonumber \\
 & & -\frac{1}{2} \sum_{ijlmm'\sigma\sigma'}
U_{mm',nn'}^{\sigma\sigma'}(ijl)
c^+_{im' \sigma}c_{jm \sigma} c^+_{ln' \sigma'}c_{ln \sigma'},
\label{eq5}
\end{eqnarray}
where $U_{mm'}^{\sigma\sigma'}(ij)$ (for $i \neq j$) describe an
attraction
between
electrons
on the nearest sites with spins $\sigma$ and $\sigma'$ and in orbitals
$m$ and $m'$ while $U_{mm',oo'}^{\sigma\sigma'}(ijl)$ ($i \neq j
\neq l$) constant is the
interaction (of any sign) between three nearest neighbour sites
respectively
(Fig.~\ref{fig2}). It is  called the assisted hopping term \cite{Cox99}. In
our
model, as will be disscused later, we
assume that this term couples the in- and out-of- plane order parameters.

The actual calculations consist of solving, self-consistently,
the following Bogoliubov-de Gennes equation:

\begin{equation}
 \sum_{jm'\sigma'} \left(\begin{array}{c}
 E^\nu - H_{mm'}(ij)  ~ ~  ~
 \Delta^{\sigma\sigma'}_{mm'} (ij)\\
 \Delta^{\sigma\sigma'*}_{mm'}(ij) ~ ~ ~ ~
 E^\nu +  H_{mm'}(ij)
\end{array}\right)
\left(\begin{array}{ll}
 u^\nu_{j m'\sigma'}\\
v^\nu_{jm'\sigma'}\end{array}\right)=0\,, \label{eq6}
\end{equation}
where  $ H_{mm'}(ij) $
is the normal spin independent part of the Hamiltonian, and
the $\Delta^{\sigma\sigma'}_{mm'}(ij)$ is   
self consistently given
in terms of the pairing amplitude, or order parameter,
$\Delta_{mm'}^{\sigma\sigma'}(ij)$ is defined by the usual relation:
\begin{eqnarray}
\Delta^{\sigma\sigma'}_{mm'}(ij) &=&  U_{mm'}^{\sigma\sigma'}(ij)
\chi_{mm'}^{\sigma\sigma'}(ij) \nonumber \\
&+&  \sum_{l oo'} 
U_{mm',oo'}(ijl)
\chi_{oo'}^{\sigma\sigma'}(il)\,,
\label{eq7}
\end{eqnarray}
where
\begin{equation}
\chi_{mm'}^{\sigma\sigma'}(ij) =
\sum_{\nu} u^\nu_{im\sigma}v^{\nu*}_{jm'\sigma'}
(1 - 2f(E^\nu))\,,
\label{eq8}
\end{equation}
and 
\begin{equation}
f(E^\nu)=\frac{1}{1+{\rm e}^{\beta E^\nu}}
\end{equation}
 is  Fermi function,
$\beta=1/k_BT$, $k_B$ is Boltzmann constant and $\nu$ enumerates the
solutions of
Eq.~\ref{eq6}.

\section{Symmetry of order parameters}  

We solved the above system of Bogoliubov de Gennes equations
(\ref{eq6}-\ref{eq8}) including all three bands and the
experimental three dimensional Fermi surface (Fig.~\ref{fig1}).
We assumed that the pairing interaction $U_{mm'}^{\sigma\sigma'}(ij)$
for nearest neighbours in plane is only acting
for the $c$  ($d_{xy}$) Ru orbitals and 
the nearest neighbour inter plane
interaction
acts only in $a$ and $b$ orbitals  ($d_{xz}$, $d_{yz}$).
The motivation for this is that the dominant hopping integrals
in plane are between $c$ orbitals, and the largest out of
plane hopping integrals are for $a$ and $b$.
On the other hand the 
additional three point, assisted hopping,
interaction
provides coupling between different orbitals a,b and c    

Therefore we have only three coupling constants
$U_\parallel$ and $U_\perp$ describing these
physically different interactions 
in- and off-plane and
$U_I$  which correspond to inter-orbital coupling. 
Our strategy is to adjust 
these phenomenological parameters
in order to obtain one transition at
the experimentally determined $T_c$.
Of course, this can be done for many choices of interactions parameters.
Compare the results of our previous orbital model \cite{Ann01} with these
obtained here for generalized orbital model with three point interactions
taken into account.

\begin{figure}[bh]
\vspace*{-0.3cm}
\hspace{0.4cm} \epsfig{file=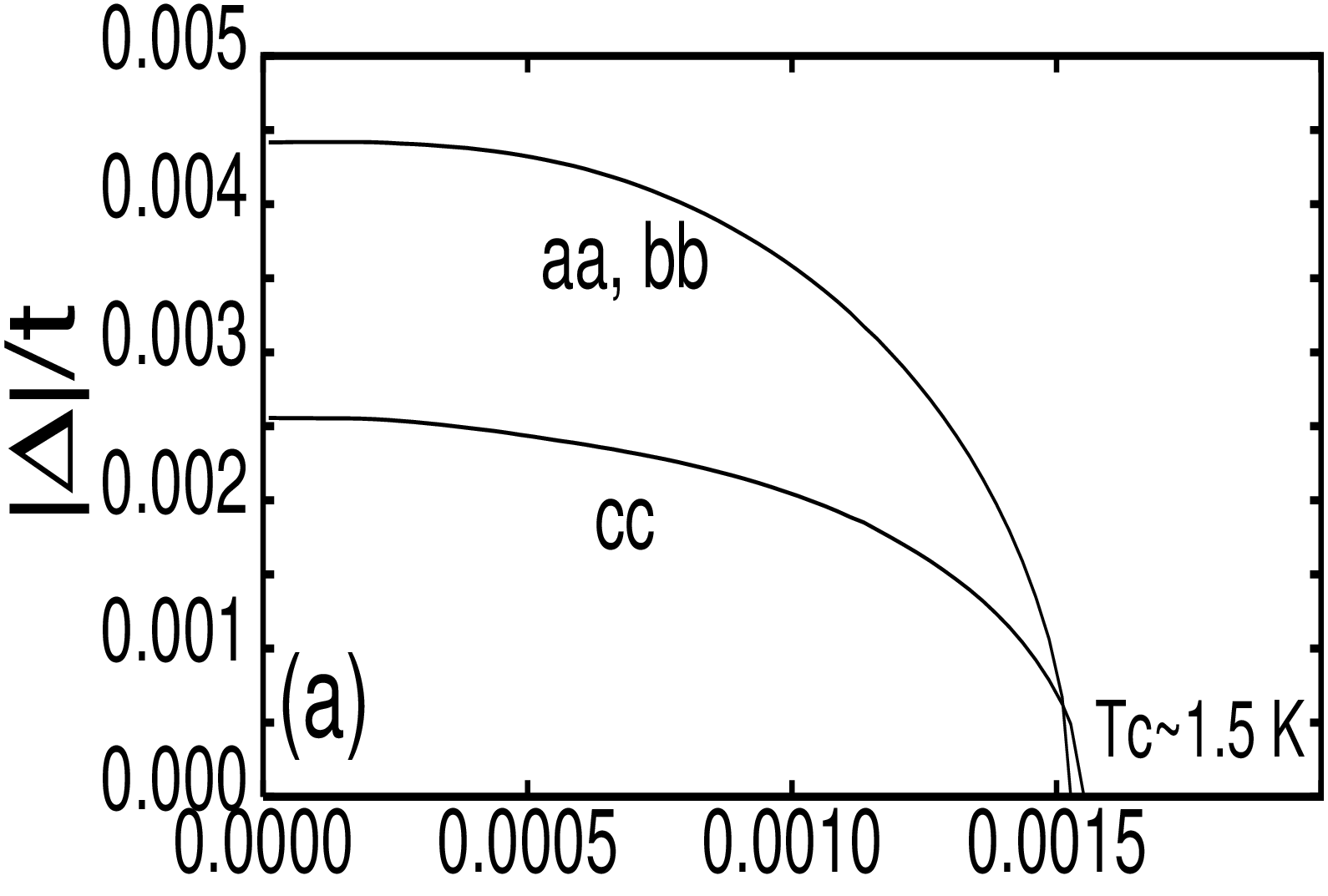,width=4.4cm,angle=-90}
\hspace{-0.915cm} \epsfig{file=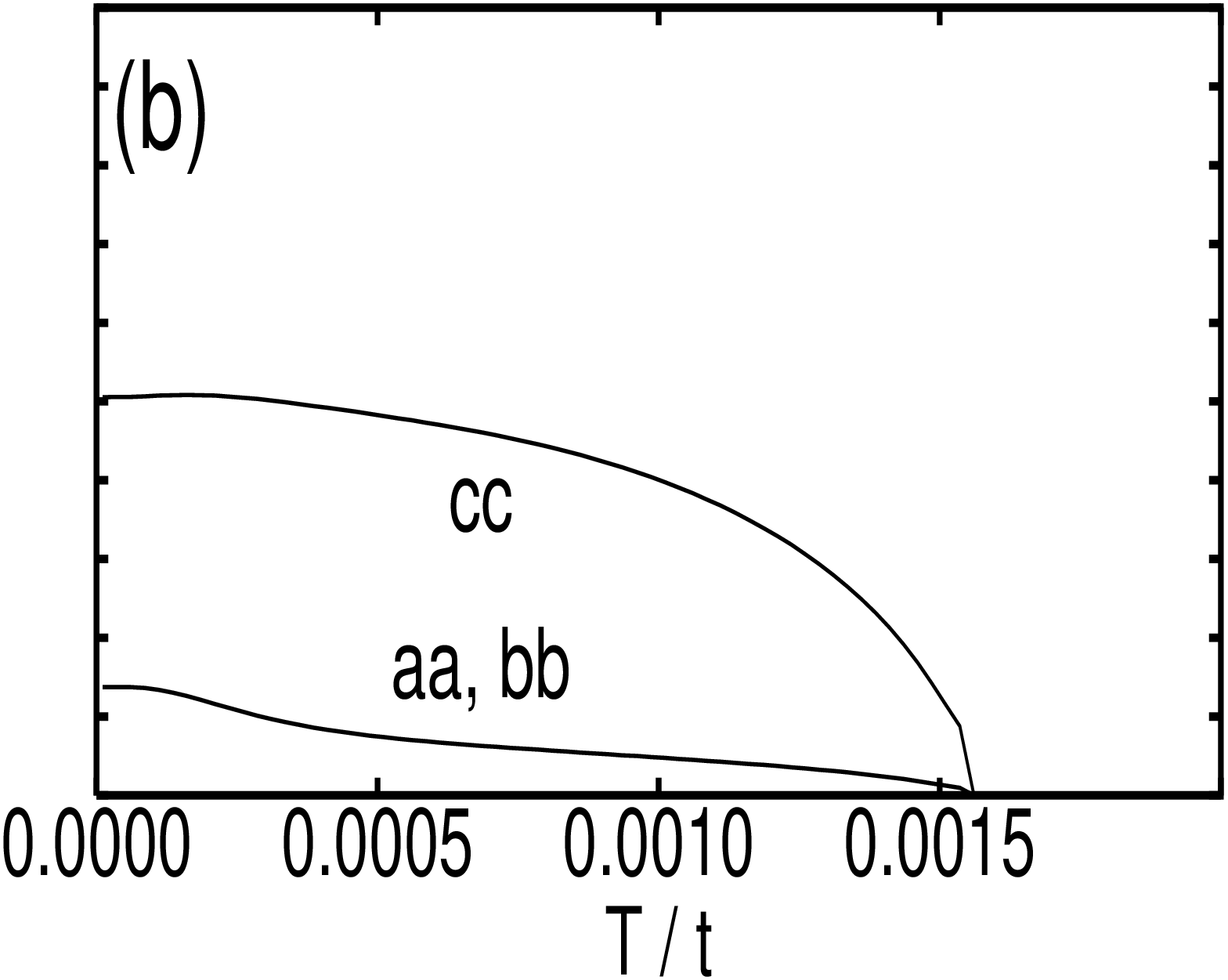,width=4.4cm,angle=-90}
\hspace{-0.915cm} \epsfig{file=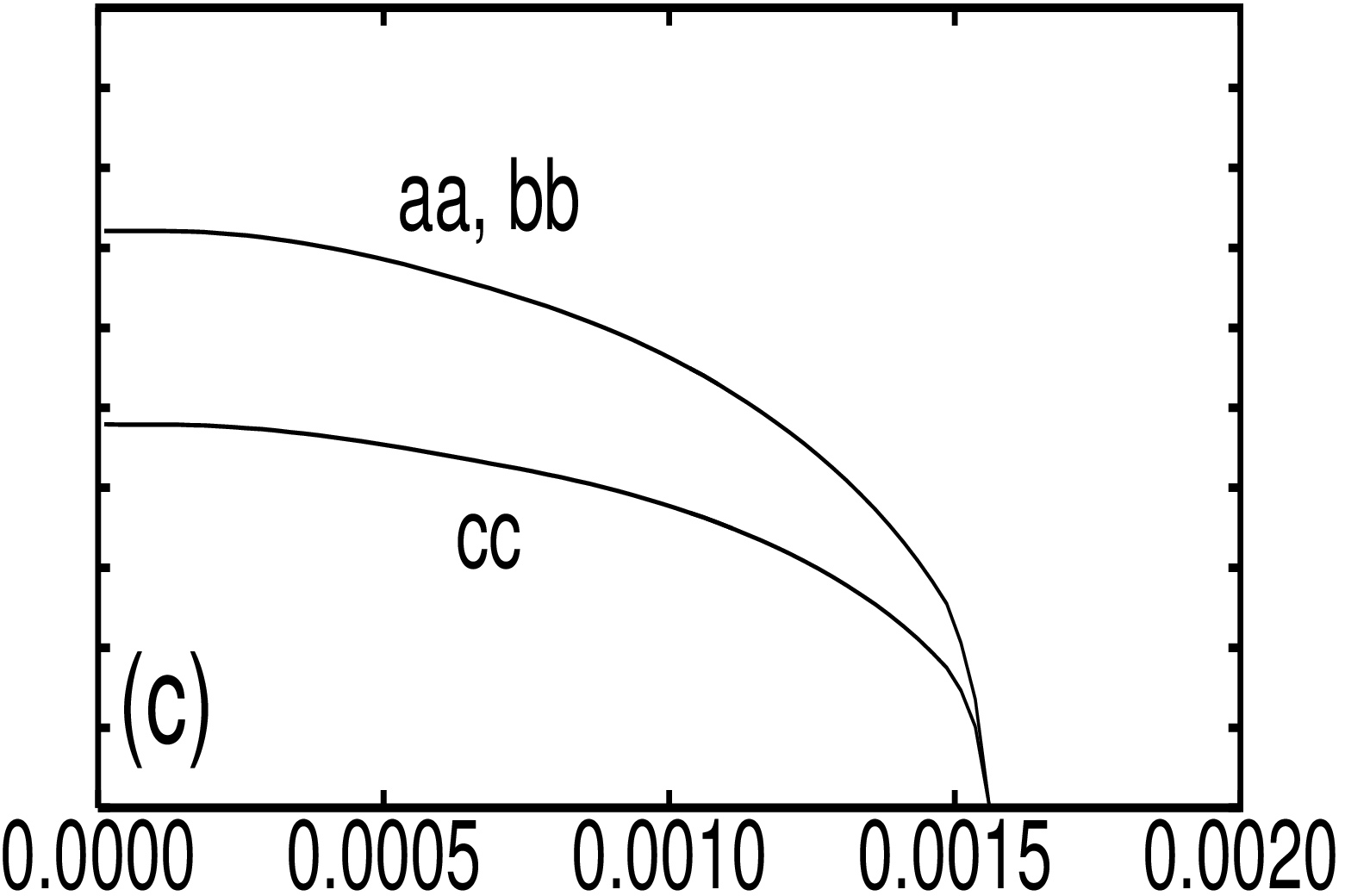,width=4.4cm,angle=-90}
\vspace{1.5cm}
\caption{
Order parameters, $|\Delta^x_{aa}|=|\Delta^x_{bb}|$ and
$|\Delta^{x}_{cc}|$
as functions of temperature for various sets of interactions: $U_{\perp}$,
$U_{\parallel}$, $U_{I}$ ((a) for $0.590t$, 
 $0.494t$, $0.0$; (b) for $0.400$, $0.493t$, $0.004t$; 
(c) for $0.400t$, $0.400t$, $0.054t$; respectively; where
$t=0.0816$ eV).}
\label{fig3}
\end{figure}

Because the pairing interactions
 $U^{\sigma\sigma'}_{mm'}(ij)$ were assumed to act
only for nearest neighbour sites in or out of plane,
the pairing potential $ \Delta^{\sigma\sigma'}_{mm'}(ij)$
is also restricted to nearest neighbours.
We further focus on only odd parity (spin triplet)
pairing states for which the vector ${\bf d} \sim (0,0,d^z)$,
i.e. $ \Delta^{\uparrow\downarrow}_{mm'}(ij)=
  \Delta^{\downarrow\uparrow}_{mm'}(ij)$, and
$ \Delta^{\uparrow\uparrow}_{mm'}(ij)=
\Delta^{\downarrow\downarrow}_{mm'}(ij)=0 $.
Therefore in general we have the following fourteen non-zero
order parameters  (i) for in plane bonds:
$\Delta_{cc}(\hat{\bf e}_x)$, $\Delta_{cc}(\hat{\bf e}_y)$,
and (ii) for inter-plane bonds:
$\Delta_{aa}({\bf R}_{ij})$,
$\Delta_{ab}({\bf R}_{ij})$,  $\Delta_{bb}({\bf R}_{ij})$
for ${\bf R}_{ij}=(\pm a/2, \pm a/2, \pm c/2)$.

\begin{figure}[thb]
\centerline{  \hspace{0.3cm}
\epsfig{file=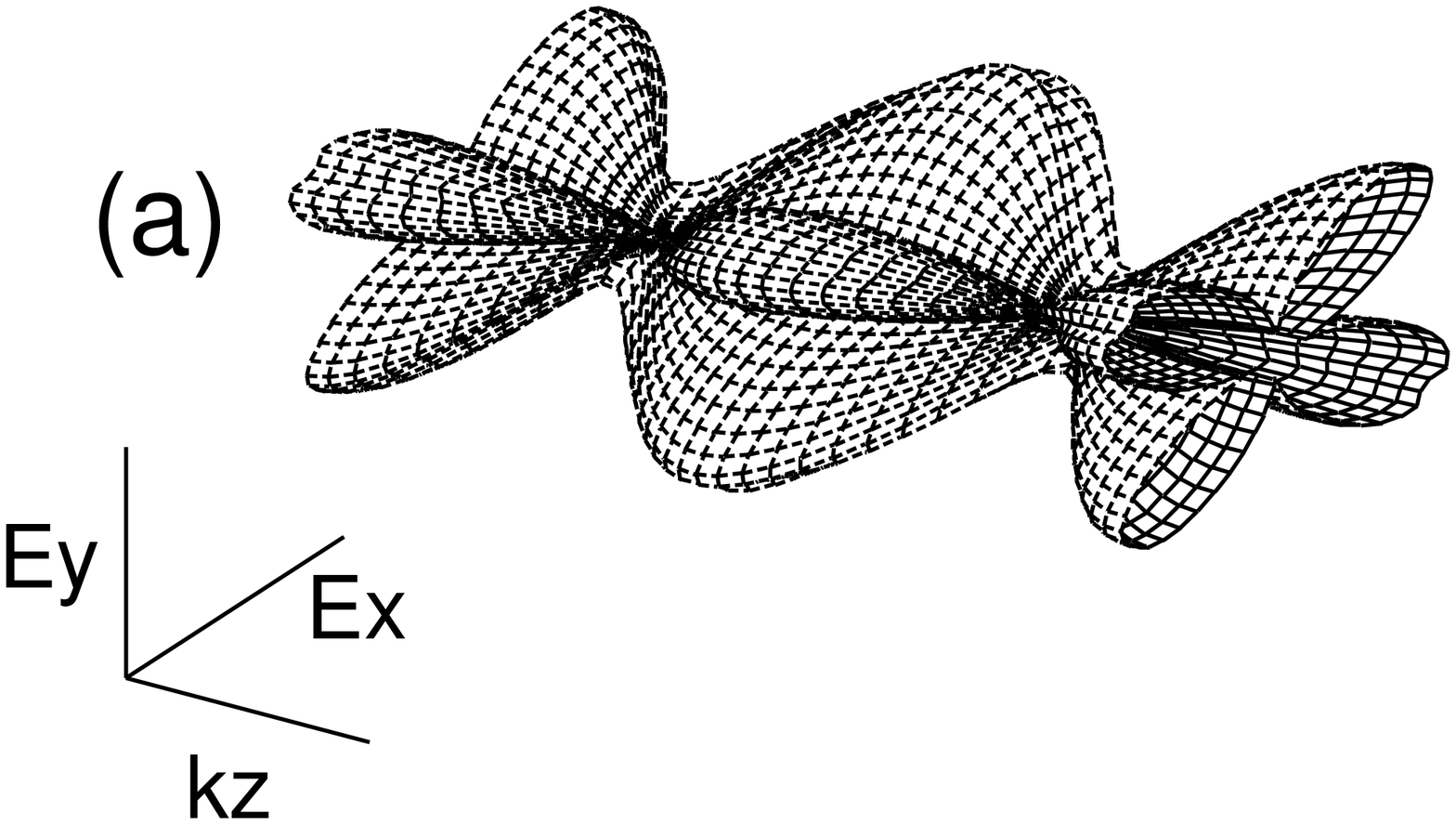,width=7cm,angle=-90} \hspace{-1.6cm}
\epsfig{file=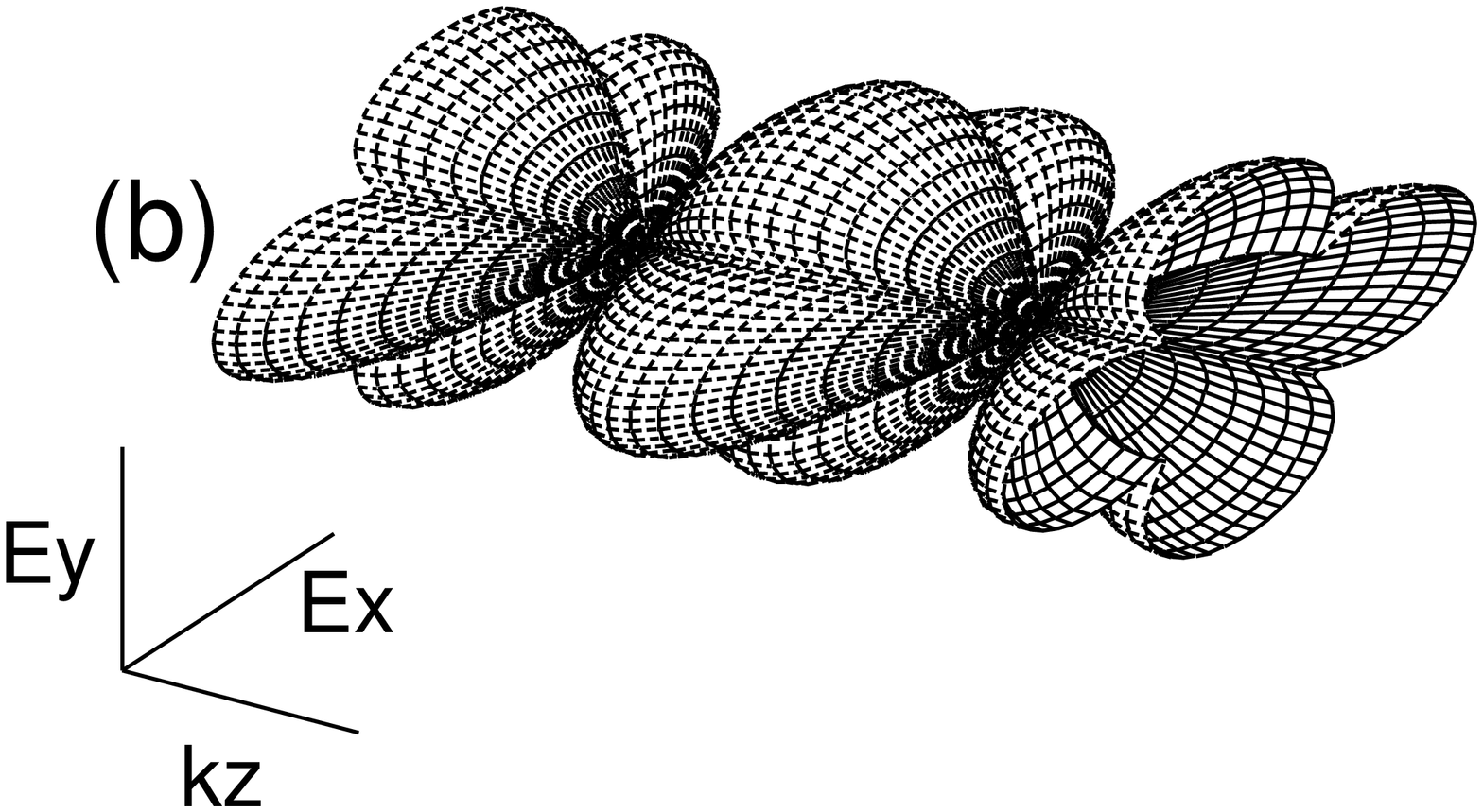,width=7cm,angle=-90}}
\vspace{-2cm}
\centerline{  \hspace{0.3cm}
\epsfig{file=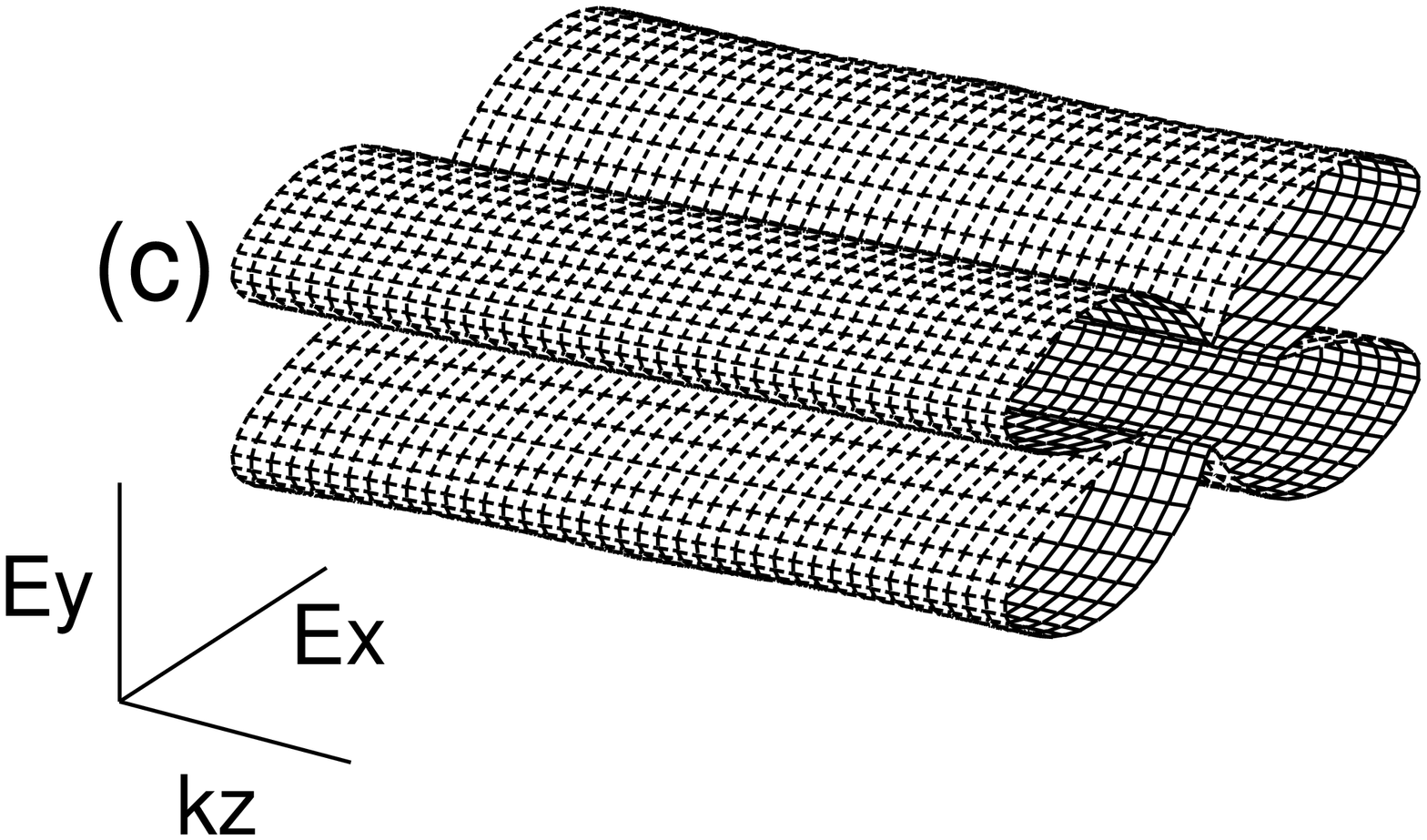,width=7cm,angle=-90} \hspace{-1.6cm}
\epsfig{file=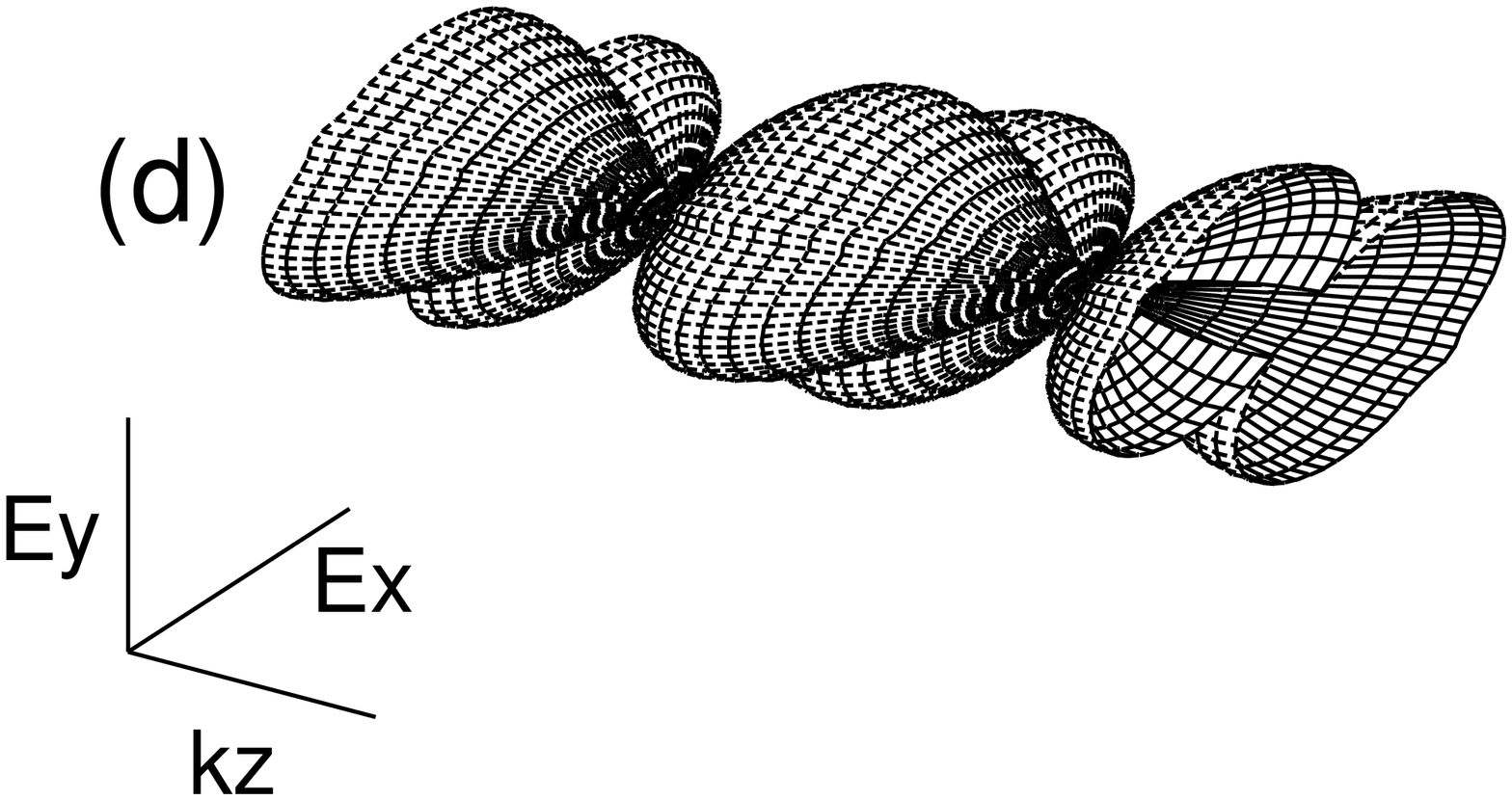,width=7cm,angle=-90}}
\vspace{-1cm}
\caption{Eigenvalues on the $\alpha$ (a), $\beta$ (b) and $\gamma$ (c) 
Fermi surface sheets, $E({\bf
k_F})$ along $k_z$ direction for interactions chosen as in
Fig.~\ref{fig3}a,
(d) $E({\bf k_F})$ for $\beta$ band Fermi 
surface and interactions chosen as in Fig.~\ref{fig3}c ($T=0$).} 
\label{fig4}
\end{figure}

Taking the lattice Fourier transform of Eq.~\ref{eq7}
the corresponding p-wave pairing potentials in k-space   
have the general form of order parameter (suppressing the spin indices
for clarity):

\begin{equation}
 \Delta_{cc}({\bf k})  = \Delta_{cc}^x  \sin{k_x} +
 \Delta_{cc}^y  \sin{k_y} 
\label{eq10}
\end{equation}
for the $c$ orbitals and,
\begin{equation}
 \Delta_{mm'}({\bf k}) 
= \left(\Delta^{x}_{mm'} \sin{\frac{k_x}{2}} \cos{\frac{k_y}{2}}
+\Delta^{y}_{mm'} \sin{\frac{k_y}{2}} \cos{\frac{k_y}{2}} \right)
\cos{\frac{k_zc}{2}} \nonumber
\label{eq11}
 \end{equation}
for $m,m'=a,b$.
A more complete k-space analysis can be found in Appendix A.

Note that beyond the usual
p-wave symmetry of the $\sin{k_x}$ and $\sin{k_y}$ type
for the $c$ orbitals, we include all three 
additional p-wave symmetries of the $\sin{k_i/2}$ type which are
induced by the effective attractive interactions between carriers on the
neighboring out-of-plane Ru orbitals.
These interactions are also responsible
for the f- and p$_z$-wave  symmetry order parameters \cite{Ann01}.
These relatively small components with smaller $T_c$ will be neglected in
this paper. The motivation comes from the strong impurity effects in the
material \cite{Mac98}.

We have solved  Eqs.~\ref{eq6}-\ref{eq8} for various sets of
interactions: $U_{\perp}$,
$U_{\parallel}$, $U_{I}$.   
For $U_I=0$ the order parameters have the symmetries
$\Delta_{cc}^y=i\Delta_{cc}^x$,
$\Delta_{bb}^y=i\Delta_{aa}^x$
as expected for a pairing symmetry \cite{Agt97}
of $(k_x+ik_y)\hat{\bf e}_z$ type
corresponding to the same time reversal broken pairing
state as $^3$He-A.
The off-diagonal components, such as
$\Delta^x_{ab}$ are small but non-zero,
as are $\Delta_{bb}^x$ and $\Delta_{aa}^y$.
Note that the k-space pairing potentials
$\Delta_{mm'}({\bf k})$
do not directly correspond to the energy gaps on the Fermi surface
sheets,
because the tight-binding Hamiltonian is non-diagonal in the orbital
indices. Instead in Fig~\ref{fig4}a-c we show the energy
eigenvalues $E_{\bf
k}$ of our Hamiltonian evaluated at the Fermi surface for $\alpha$,
$\beta$ and $\gamma$ bands. One can see
 that there  are horizontal
circles around the cylindrical
$\beta$ Fermi surface sheet (Fig.~\ref{fig4}b), 
while the order parameter at the $\gamma$ sheet  (Fig.~\ref{fig4}c) is
node-less. Interestingly,
in our model
$\alpha$ sheet does not posses line nodes.  Due to the small diagonal   
distortion of $\alpha$ Fermi surface  (Fig.~\ref{fig1}) it has point
nodes  (Fig.~\ref{fig4}a).

$U_{\perp}$,
$U_{\parallel}$, $U_{I}$ were chosen in such way as to give a single
transition with $T_c=1.5$ K.
Figure~\ref{fig3}a shows various $\Delta$ obtained for our
\cite{Ann01} orbital model  ($U_{\perp}=0.590t$,
 $U_{\parallel}=0.494t$ but  $U_{I}=0.0$). Here the interactions were tuned 
separately to give approximately the same $T_c$. All components of the  order
parameter in a,b and c orbitals are active.

\begin{figure}[tbh]
\vspace{-0.3cm}
\centerline{
\hspace{-0.655cm} \epsfig{file=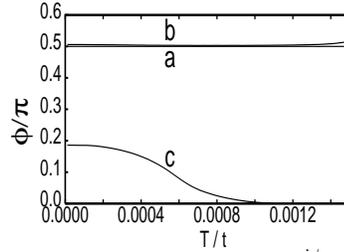,width=4.3cm,angle=-90}}
\vspace{1.5cm}
\caption{Temperature dependence of $\phi$ 
($\Delta_{cc}^y = {\rm e}^{{\rm i} \phi} \Delta_{cc}^x $). 
Interaction parameters $U_{\perp}$,
$U_{\parallel}$, $U_{I}$ as in Fig. \ref{fig3} for (a), (b) and (c);
respectively.
Note that for  $U_{I}=0$  (a) $\phi=\pi/2$ and $\Delta_{cc}^y = {\rm i}
\Delta_{cc}^x $. }      
\label{fig5}
\end{figure}   

In the case of Fig.~\ref{fig3}b
$U_{\perp}$,
$U_{\parallel}$, $U_{I}$ are equal to
$0.400t$, $0.493t$, $0.004t$, respectively. 
Here  a,b orbital  order parameters $\Delta^x_{aa}$, $\Delta^x_{bb}$
have
much smaller values than $\Delta^x_{cc}$ and only due to
small three point inter orbital coupling  $U_{I}$. We have the proximity
effect
and hence a
unique $T_c$. 

In the last case
$U_{\perp}$=$0.400t$, $U_{\parallel}=0.400t$, $U_{I}=0.054t$
 (Fig.~\ref{fig3}c)
the situation is different. To get $T_c=1.5K$ both interactions
$U_{\perp}$,
$U_{\parallel}$ take on  smaller values comparing to those in
Fig.~\ref{fig3}a.
The increase of the inter-orbital interaction $U_I$ changes $T_c$ and the
values of  $\Delta_{mm'}^i(T=0)$ as well as their temperature dependence.
The order parameters $\Delta_{mm}^i$
fulfill more general relations $\Delta_{cc}^x = {\rm e}^{{\rm i} \phi}
\Delta_{cc}^y $ and $\Delta_{bb}^x = {\rm e}^{{\rm i} \phi}    
\Delta_{aa}^y$, where $\phi$ is a temperature dependent phase. This
dependence for all three cases are plotted in Fig.~\ref{fig5}. 
For $U_I=0$ $\phi$ is constant and  equal to $\pi/2$ (curve a). A similar
behaviour with $\phi \approx \pi/2$ can be found for small $U_I$ (curve b)
while
for relatively large $U_I$ ($U_I=0.054t$) $\phi$ depends strongly on
temperature (curve c).  
It is worthwhile to note that for low temperature $T$ all order
parameters  are
complex. It is also true for higher $T$ but for  $U_I$ equal to 0 or
relatively small. 
For larger interaction $U_I$ there is a
temperature $T^*$ ($T^* < T_c$) above which the order parameters are real.
This is a source of  additional vertical line nodes appearing at the Fermi
surface.
For $U_I=0.054t$  the horizontal line nodes are still
present on the
 cylindrical $\beta$ Fermi surface sheet (Fig.~\ref{fig4}d). Due to $\phi
\neq \pi/2$ the eigenvalues on the Fermi surface
show a two-fold
symmetry instead of four-fold one present for $U_I=0$ (Fig.~\ref{fig4}b).

\begin{figure}[tbh]
\vspace{-0.3cm}
\hspace{0.4cm} \epsfig{file=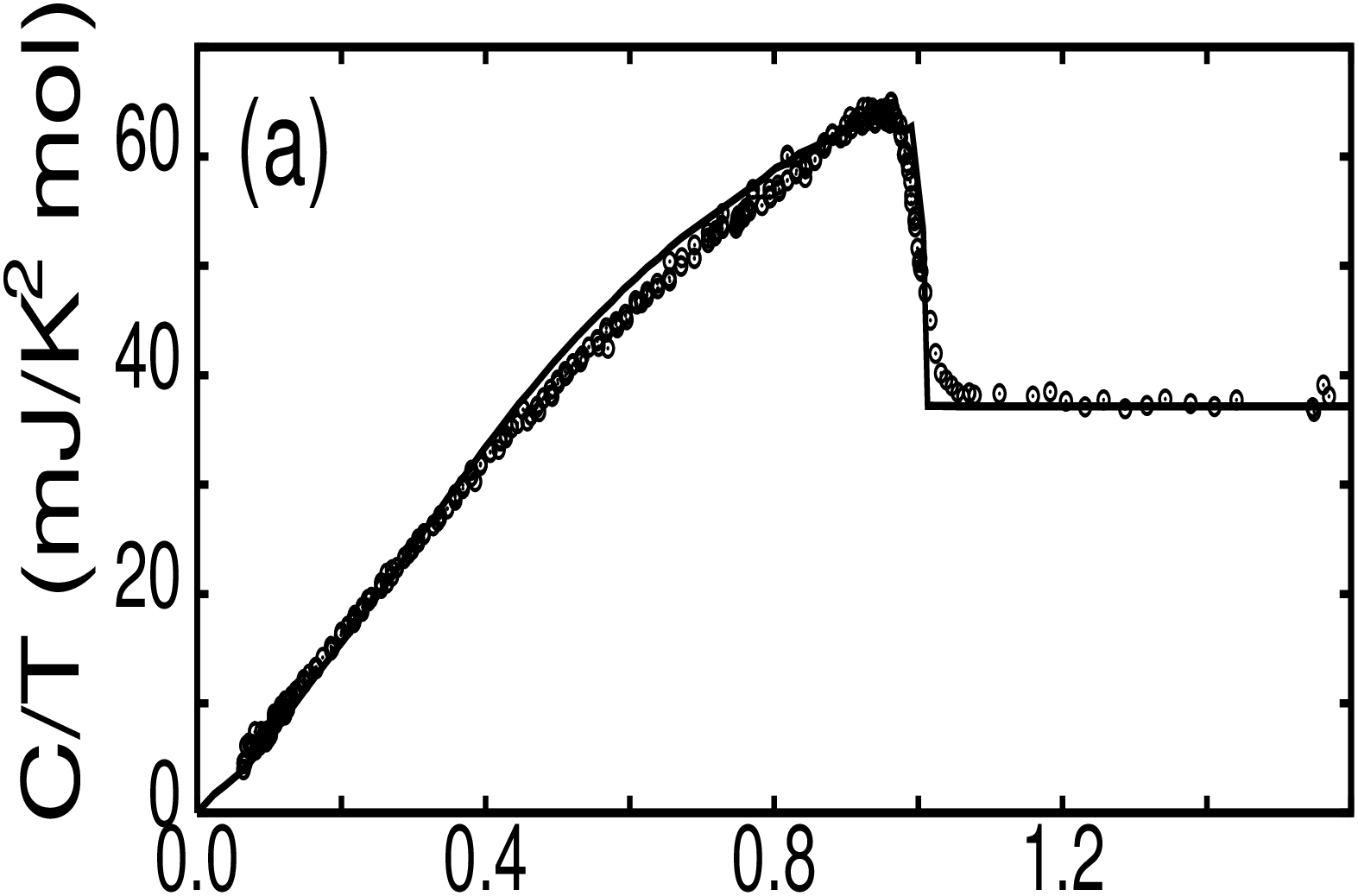,width=4.3cm,angle=-90}
\hspace{-0.655cm} \epsfig{file=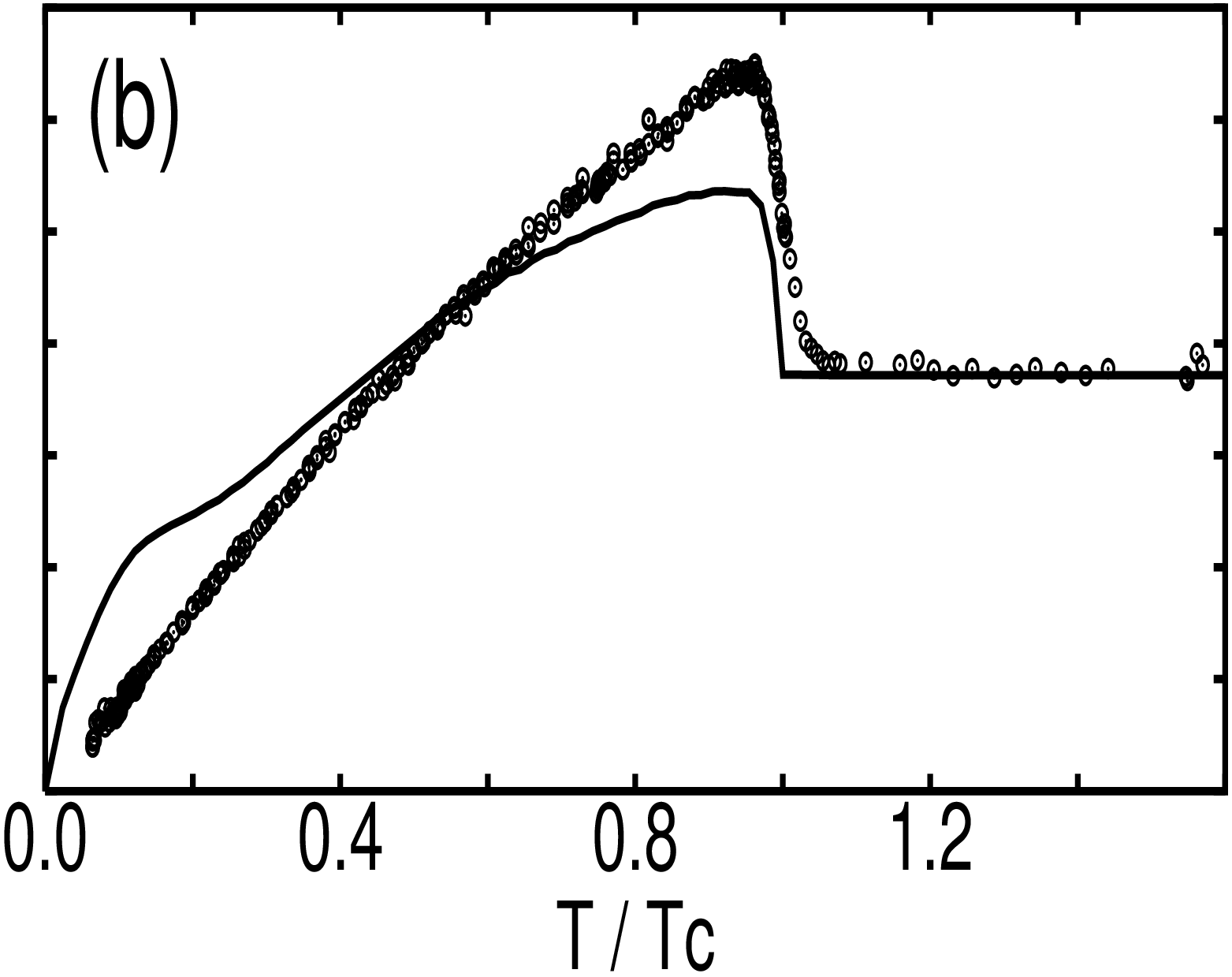,width=4.3cm,angle=-90}
\hspace{-0.655cm} \epsfig{file=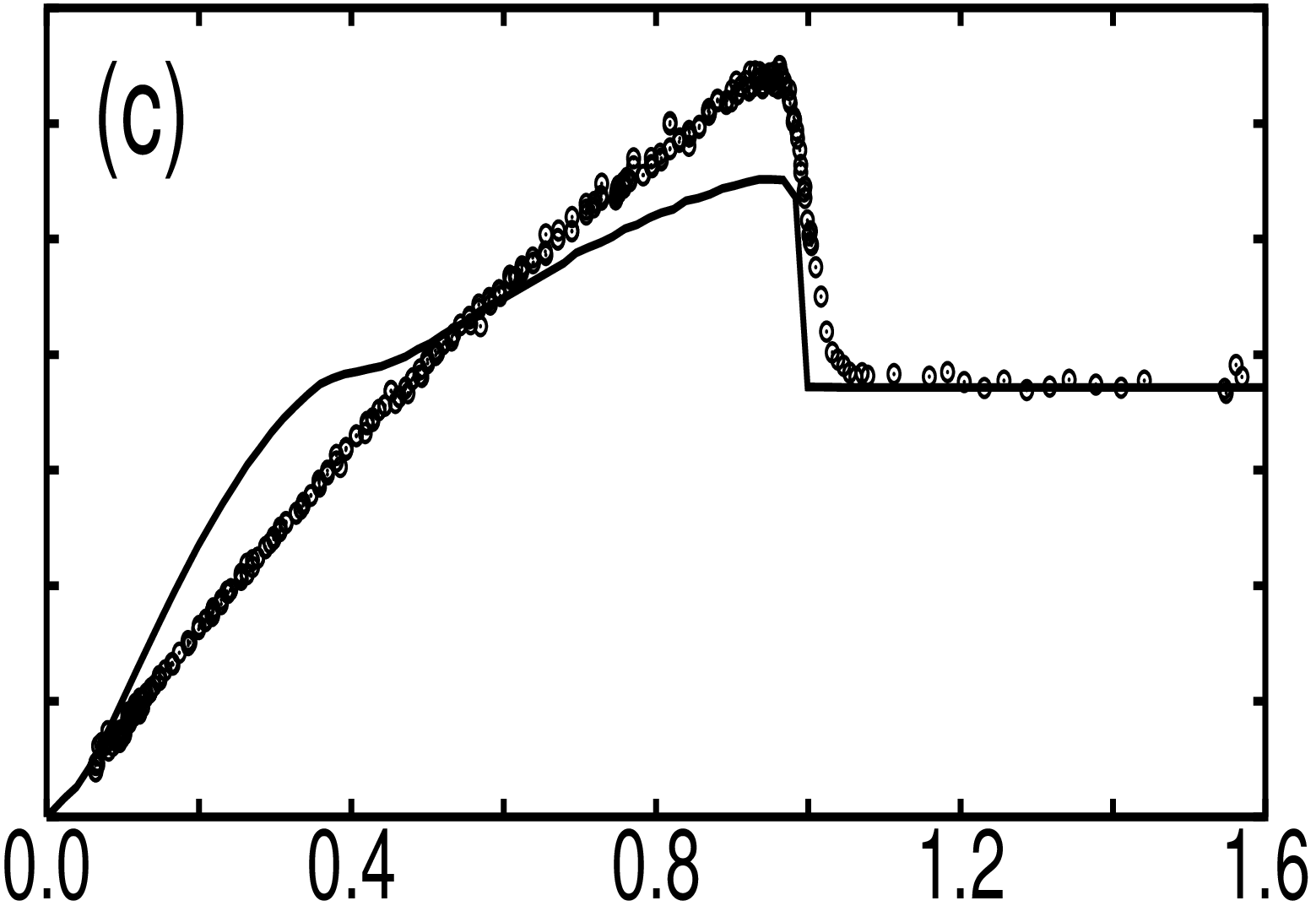,width=4.3cm,angle=-90}
\vspace{1.0cm}
\caption{Lines: Specific heat calculated for various sets of interactions
(parameters $U_{\perp}$,
$U_{\parallel}$, $U_{I}$ as in Fig.~\ref{fig3});
full circles denote experimental data [16]}
\label{fig6}
\end{figure}

To decide which of these three cases is closer to experiment
we calculated the specific heat  for the same  interaction parameters as
above using the relation: 
\begin{equation}
C = -2  k_B \beta^2 \sum_{m, \nu} E_{m, \nu} \frac{
\partial f (E_{m, \nu}) }{\partial \beta} \,.  
\label{eq12}
\end{equation}

In Figs.~\ref{fig6} a--c we have plotted
the specific heat versus temperature.
In all cases the low temperature limit of the specific heat is power
law, because our gap parameters have line nodes on $\beta$ Fermi surface
sheet.

Going to larger temperatures we observe 
in Fig.~\ref{fig6}a, that $C$
has a discontinuity at $T_c$ of $27$mJ/mol, which is in
agreement with the experiment \cite{Nis00,Ann01}. The slope of
$C/T$ versus $T$ also comes out correctly.
The values of the interaction parameters $U_{\perp}$ and  $U_{\parallel}$
are in this model tuned to give single superconducting transition at
temperature equal to 1.5 K.
Interestingly, the two other figures (Fig.~\ref{fig6}b-c)  which
correspond to the inter-orbital proximities have too small value of the
specific heat jump at $T_c$. Technically it is due to different
curvatures of temperatures dependences of $\Delta_{mm'}$.

\section{Summary and conclusions}

In summary we would like to emphasize two points. Firstly, we have
analyzed the orbital model of superconductivity with and without
 `inter-orbital proximity effect'. Both versions of the model allow for
complex order parameter with line nodes on some of the Fermi surface
sheets.
 Our
description is a real space one with  two point tight-binding interaction
such as
naturally arises in multi-band, extended, negative $U$ Hubbard model
\cite{Mic91}.

Secondly, we wish to stress that in our original approach to the problem
the
parameters which describe the normal state are determined by fitting
them to
the
very accurately known Fermi surface.  The measured $T_{c}$ determines
both  coupling constants $U_\parallel$,
$U_\perp$.
As for the physical mechanism of pairing, the fact that
$U_\perp \approx U_\parallel$ implies that the pairing interaction is
fairly isotropic in spite of the layered structure of the system. The
model with "inter-orbital proximity" effect requires even more fine tuning
of parameters in order to get correct slope and jump of the specific heat. 

In conclusion, the scenario with active cc orbitals and "orbital proximity"
mechanism for superconductivity in a and b orbitals is not consistent with
specific heat data on Sr$_2$RuO$_4$.

\section{Acknowledgements}
This work was partially founded by the Committee of Scientific Research (Poland) 
through the grant KBN 2P03B 106 18, and the Royal Society (UK).

\section{Appendix A}
\setcounter{equation}{0}
\def\theequation{A.\arabic{equation}}  

In case of a pure system one can take a Fourier transform of
Eqs.~(\ref{eq6}-\ref{eq8}).
Thus
\begin{equation}
 \sum_{m'\sigma'} \left(\begin{array}{c}
 E_{\bf k}^m - H_{mm'}({\bf k})  ~ ~  ~
 \Delta^{\sigma\sigma'}_{mm'} ({\bf k})\\
 \Delta^{\sigma\sigma'*}_{mm'}({\bf k}) ~ ~ ~ ~
 E^m_{\bf k} +  H_{mm'}({\bf k})   
\end{array}\right)
\left(\begin{array}{ll}
 u^\nu_{{\bf k} m'\sigma'}\\
v^\nu_{ {\bf k} m'\sigma'}\end{array}\right)=0\,, 
\label{eqa1}
\end{equation}

The gap equation (Eqs.~\ref{eq7}-\ref{eq8}) can be rewritten in k-space 
as

\begin{eqnarray}
\Delta^{\sigma\sigma'}_{mm'}({\bf k}) &=&
\frac{1}{N}\sum_{\bf q} U_{mm'}^{\sigma\sigma'}({\bf k}- {\bf q})
\chi_{mm'}^{\sigma\sigma'}({\bf q}) \nonumber \\
&+&  \frac{1}{N} \sum_{{\bf q}, oo'}
U_{mm',oo'}({\bf q},{\bf k} -{\bf q})
\chi_{oo'}^{\sigma\sigma'}({\bf q})\,.
\label{eqa2}
\end{eqnarray}
and
\begin{equation}
\chi_{mm'}^{\sigma\sigma'}({\bf k}) =
u_{{\bf k}m\sigma}v^*_{{\bf k}m'\sigma'}
(1 - 2f(E_{\bf k}^m))\,.
\label{eqa3}
\end{equation}

In the orthogonal crystal (Fig.~\ref{fig2}) various matrix elements of the
interaction $U$ responsible for p-wave paring can be
written as

\begin{eqnarray}
U_{cc}({\bf k},{\bf q}) &=& 2 U_{\parallel} V({\bf k})  V({\bf q})
\nonumber \\  
 U_{mm'}({\bf k},{\bf q}) &=& 8 U_{\perp} \tilde V({\bf k})  \tilde
V({\bf q}) ~~~~~~~~~~~{\rm for}~~ m,m'=a,b \label{eqa4} \\
 U_{mm'cc}({\bf k},{\bf k}') &=& 8 U_{I} \tilde V({\bf k}) V({\bf
q}) ~~~~~~~~~~~{\rm for}~~ m',m=a,b \nonumber \\
U_{ccmm'}({\bf k},{\bf q}) &=& 8 U_{I} V({\bf k})  \tilde V({\bf
q}) ~~~~~~~~~~~{\rm for}~~ m',m=a,b \nonumber
\end{eqnarray}
where $V({\bf k})$ and $\tilde V({\bf k})$ can be expressed as
\begin{eqnarray}
&& V({\bf k})= \left( \sin{k_x} +  \sin{k_y}\right) 
\label{eqa5} 
\\
&& \tilde V({\bf k})= \left(\sin{\frac{k_x}{2}}
\cos{\frac{k_y}{2}} +  \sin{\frac{k_y}{2}}
\cos{\frac{k_y}{2}} \right) \cos{\frac{k_zc}{2}} \, \nonumber.
\end{eqnarray}

Thus, the general form of order parameter

\begin{equation}
 \Delta_{cc}({\bf k})  = \Delta_{cc}^x  \sin{k_x} +
 \Delta_{cc}^y  \sin{k_y}
\label{eq10b}  
\end{equation}
for $c$ orbitals and,
\begin{equation}
 \Delta_{mm'}({\bf k})
= \left(\Delta^{x}_{mm'} \sin{\frac{k_x}{2}} \cos{\frac{k_y}{2}} 
+\Delta^{y}_{mm'} \sin{\frac{k_y}{2}} \cos{\frac{k_y}{2}} \right)
\cos{\frac{k_zc}{2}} \nonumber
\label{eq11b}   
 \end{equation}
for $m,m'=a,b$ orbitals.


\begin{thebibliography}{99}
\bibitem{Leg75} A. J. Leggett, Rev. Mod. Phys, {\bf 47} 331 (1975).
\bibitem{Mae01} Y.Maeno, T.M. Rice and M. Sigrist,
 Physics Today {\bf 54}, 42 (2001).
\bibitem{Mac00} A. Mackenzie and Y. Maeno, Physica {\bf B280}, 148
(2000).
\bibitem{Agt97} D. F. Agterberg, T. M. Rice and M. Sigrist, Phys. Rev.
Lett. {\bf 73} 3374 (1997).
\bibitem{Maz97} I. I. Mazin and D. J. Singh, Phys. Rev. Lett. {\bf 79}
 733 (1997).
\bibitem{Miy99} K. Miyake and D. Narikiyo, Phys. Rev. Lett. {\bf 83}
(1999).
\bibitem{Gra00} M.J. Graf and A.V. Balatsky, Phys. Rev. b {\bf 62},
9697 (2000).
\bibitem{Won00} H. Won and K. Maki, Europhys. Lett. {\bf 52},  427 (2000).
\bibitem{Dah00} T. Dahm, H. Won, and K. Maki, cond-mat/0006301.
\bibitem{Sig00} M. Sigrist, Physica {\bf B280}  154 (2000).
\bibitem{Has00} Y. Hasegawa, K. Machida and M. Ozaki, J. Phys. Japan
{\bf 69}, 336 (2000).
\bibitem{Zhi01} M. E. Zhitomirsky and T. M. Rice, Phys. Rev. Lett. {\bf
87},
 057001 (2001).
\bibitem{Ann01} J.F. Annett, G. Litak, B.L. Gy\"{o}rffy, K.I. 
Wysoki\'nski,
preprint cond-mat/0109023.
\bibitem{Ere01} I. Eremin, D. Manske, C. Joas and K.H. Bennemann,
   preprint cond-mat/0102074.
\bibitem{Luk98} G.M. Luke, Y. Fudamoto,K.M. Kojima {\it et al.}
Nature {\bf 394} 558 (1998).
\bibitem{Nis00} S. NishiZaki, Y. Maeno and Z. Mao, J. Phys. Japan
{\bf 69}, 336 (2000).
\bibitem{Bon00} I. Bonalde, B.D. Yanoff, M.B. Salamon {\it et al.}, Phys.
Rev. Lett.
{\bf 85}, 4775 (2000).
\bibitem{Lup01} C. Lupien, W.A. MacFarlane, C. Proust {\it et al.},
Phys. Rev. Lett. {\bf 86}, 5986 (2001).
\bibitem{Tan01} M.A. Tanatar, M. Suzuki, S. Nagai {\it et al.},
Phys. Rev. Lett. {\bf 86} 2649 (2001).
\bibitem{Ann90} J.F. Annett, Adv. Phys. {\bf 39},  83 (1990).
\bibitem{Ber00} C. Bergemann, S. R. Julian, A.P. Mackenzie  {\it et al.},
Phys. Rev. Lett. {\bf 84} 2662 (2000). 
\bibitem{Cox99}  D.L. Cox and A. Zawadowski,
{\it Exotic Kondo Effects in Metals}, (Taylor and Francis, London 1999).
\bibitem{Mac98} A.P. Mackenzie, R.K.W. Haselwimmer, A.W. Tyler  {\it et
al.}, Phys. Rev. Lett. 80 (1998) 161.
\bibitem{Mic91} R.Micnas, J.Ranninger and S.Robaszkiewicz,
 Rev. Mod. Phys. {\bf 62}, 113 (1991).
\end{thebibliography}
\end{document}